\documentclass[twocolumn,prx,floatfix,superscriptaddress,nobibnotes]{revtex4-2}
\usepackage{graphicx}
\usepackage{enumerate}
\usepackage[colorlinks=true,citecolor=blue]{hyperref}
\usepackage{textcomp}
\usepackage{amsmath}
\usepackage{amssymb}
\usepackage{pgf}
\usepackage{subfigure}
\usepackage[english]{babel}
\usepackage[normalem]{ulem}
\usepackage{tabularray}
\usepackage{xcolor}
\usepackage{array}
\usepackage{comment}

\graphicspath{{./}{./}}

\begin{document}
\title{Tuning the Josephson diode response with an ac current}


\author{Rub\'en Seoane Souto}
\affiliation{Instituto de Ciencia de Materiales de Madrid (ICMM), Consejo Superior de Investigaciones Científicas (CSIC),
Sor Juana Inés de la Cruz 3, 28049 Madrid, Spain.}
\affiliation{Division of Solid State Physics and NanoLund, Lund University, S-22100 Lund, Sweden}
\affiliation{Center for Quantum Devices, Niels Bohr Institute, University of Copenhagen,  2100 Copenhagen, Denmark}

\author{Martin Leijnse}
\affiliation{Division of Solid State Physics and NanoLund, Lund University, S-22100 Lund, Sweden}
\affiliation{Center for Quantum Devices, Niels Bohr Institute, University of Copenhagen,  2100 Copenhagen, Denmark}

\author{Constantin Schrade}
\affiliation{Center for Quantum Devices, Niels Bohr Institute, University of Copenhagen,  2100 Copenhagen, Denmark}

\author{Marco Valentini}
\affiliation{Institute of Science and Technology Austria, Am Campus 1, 3400 Klosterneuburg, Austria.}

\author{Georgios Katsaros}
\affiliation{Institute of Science and Technology Austria, Am Campus 1, 3400 Klosterneuburg, Austria.}

\author{Jeroen Danon}
\affiliation{Department of Physics, Norwegian University of Science and Technology, NO-7491 Trondheim, Norway}

\begin{abstract}
Josephson diodes are superconducting elements that show an asymmetry in the critical current depending on the direction of the current.
Here, we theoretically explore how an alternating current bias can tune the response of such a diode.
We show that for slow driving there is always a regime where the system can only carry zero-voltage dc current in one direction, thus effectively behaving as an ideal Josephson diode.
Under fast driving, the diode efficiency is also tunable, although the ideal regime cannot be reached in this case.
We also investigate the residual dissipation
due to the time-dependent current bias and show that it remains small.
All our conclusions are solely based on the critical current asymmetry of the junction, and are thus compatible with any Josephson diode.
\end{abstract}

\maketitle

Superconductivity offers the potential for a new generation of electronic devices characterized by minimal or zero dissipation and rapid response times~\cite{Braginski_JSNM2019}. Within this promising landscape, the non-reciprocal phenomenon in superconducting systems known as the ``superconducting diode effect'' has garnered substantial attention in recent times~\cite{Margaris_JoP2010,Yokoyama_PRB2014,Silaev_JoP2014,Dolcini_PRB2015,Semenov_IEE2015,Wakatsuki_Science2017,Wakatsuki_PRL2018,Chen_PRB2018,Pal_EuL2019,ando2020observation,Mayer_NatCom2020,Lyu_NatPhys2021,Kopasov_PRB2021,Diez_arXiv21,He_NJP2022,Daido_PRL2022,Davydova_SciAdv22,Souto_PRL2022,Fominov_PRB2022,Baumgartner_NatNano2022,Pal_NatPhys2022,Lin_NatPhys2022,Tanaka_PRB2022,Yuan_PNAS2022,Bauriedl_NatCom2022,He_NJP2022,Kokkeler_PRB2022,Ilic_PRL2022,Ilic_PRA2022,Karabassov_PRB2022,Daido_PRL2022,Halterman_PRB2022,Haenel_arXiv2022,Fu_arXiv2022,Turini_NanoLett2022,Wei_PRB2022,Legg_PRB2022,Wang_arXiv2022,Song_arXiv2022,Legg_arXiv2023,Maiani_PRB2023,Bo_PRL2023,Pillet_PRR2023,Kokkeler_arXiv2023,Lotfizadeh_arXiv2023,Yasen_PRL2023,Chiles_NanoLett2023,Valentinin_arXiv2023,Karabassov_CM2023,Picoli_PRB2023,Banerjee_arXiv2023,Vigliotti_NanoMat2023,He_NatCom2023,Zhao_Science2023,Chen_AFM2023,Zazunov_arXiv2023,Lu_arXiv2023,Costa_PRB2023,Ciaccia_PRR2023,Song_ComPhys2023,Mazur_arXiv2023,Steiner_PRL2023,Trahms_Nat2023,Gupta_NatCom2023,Bozkurt_2023,Chen_arXiv2023,Cayao_arXiv2023,Liu_arXiv2023,Margineda_arXiv2023,Leblanc_arXiv2023}, for a recent review see Ref.~\cite{Nadeem_NRM2023}.

In these systems, the critical currents in the two directions are different, $|I_{c}^+|\neq |I_{c}^-|$. The conventional figure of merit for such superconducting diodes is the diode efficiency, defined by $\eta=|(I_{c}^++I_{c}^-)/(I_{c}^+-I_{c}^-)|$. This metric quantifies the asymmetry in critical currents, a pivotal aspect of diode functionality. Therefore, maximizing $\eta$ is an important aspect for potential applications of superconducting diodes. An ideal diode ($\eta=\pm 1$) is characterized by supporting supercurrent only in one direction. So far, different directions have been explored to approach unity efficiency, including multiple Andreev reflections after applying a small bias voltage~\cite{Zazunov_arXiv2023}, concatenating several junctions in parallel~\cite{Souto_PRL2022,Bozkurt_2023}, and three terminal superconducting devices (triodes)~\cite{Chiles_NanoLett2023}. Very recently, there was a proposal for an ideal diode with dissipation based on the application of an electric field perpendicular to the supercurrent propagation~\cite{Daido_arXiv2023}.

On the other hand, Ref.~\cite{Valentinin_arXiv2023} demonstrated experimentally how a periodic modulation of the bias current, $I_{\rm ac}$, can tune the effective diode efficiency. Even though the system had a relatively low $\eta$ for $I_{\rm ac}=0$, it could be tuned to the ideal regime at finite $I_{\rm ac}$, with a measurable zero-resistance plateau that only extends for one direction of the dc bias current. The possibility of modulating the diode response using time-dependent voltage biasing has been studied theoretically in Ref.~\cite{Cuozzo_arXiv2023} in topological junctions, demonstrating a regime where $\eta$ effectively approaches one, but with very small critical current. The ideal regime was also analyzed theoretically in quantum dot systems subject to two ac signals out of phase~\cite{Ortega_PRB2023,Soori_PS2023}.

\begin{figure}[t]
\begin{center}
    \includegraphics[width=1.0\linewidth]{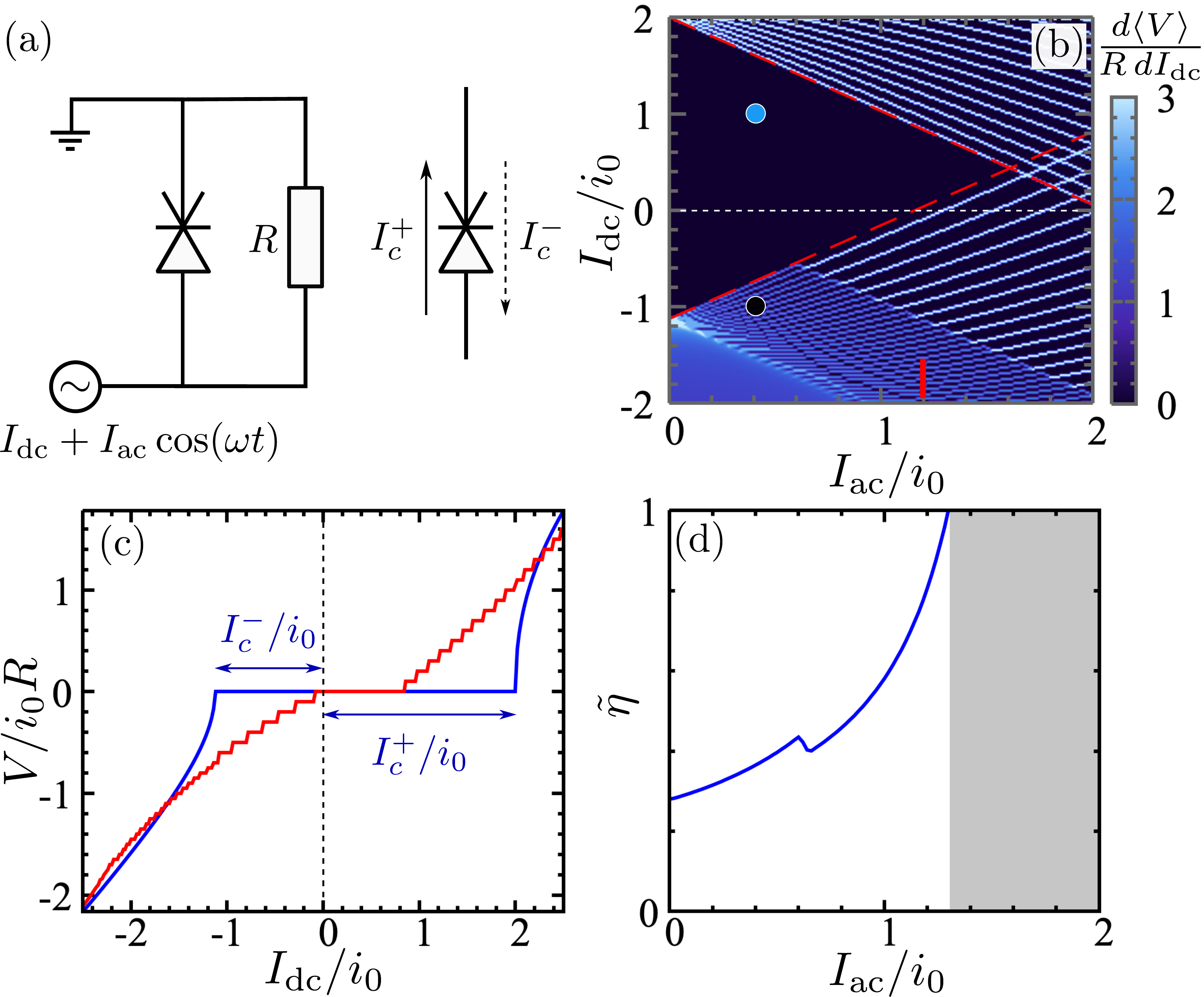}
\end{center}
\caption{{\bf Diode response in the slow-driving regime.}
(a) Schematic of the circuit we consider.
(b) Average differential resistance as a function of the two current biases. Red dashed lines show the boundaries of the zero-voltage region as predicted by the adiabatic theory, given by $\tilde I^{\pm}_{\rm dc}=I^{\pm}_{c}\mp I_{\rm ac}$. The black and blue dots denote the parameters used for the lower panels of Fig.~\ref{fig:potentials}.
(c) Average voltage drop across the junction as a function of $I_{\rm dc}$, for $I_{\rm ac}=0$ (blue) and $I_{\rm ac}=1.2i_0$ (red).
(d) Effective diode efficiency as a function of $I_{\rm ac}$, as extracted from (b). In  all panels, we used $I(\varphi)=i_0\sin(\varphi)+i_0\sin(2\varphi-\frac{\pi}{2})$ and $\omega_0=0.1i_0$.
\label{fig1}}
\end{figure}

Motivated by the recent experimental observations of Ref.~\cite{Valentinin_arXiv2023}, we present here the theoretical analysis of the current-driven superconducting diode. We consider a superconducting diode subject to a current bias that has a dc and an ac component, see Fig.~\ref{fig1}(a). We focus on two limiting scenarios where the frequency of the ac driving is either much smaller or much larger than the inverse characteristic time of the junction, related to its critical current and normal resistance, and we derive analytical insight in the diode response in both cases. We explain why, independent of the origin of the diode effect, a slowly varying $I_{\rm ac}$ can tune the diode to become effectively ideal. In the opposite fast-driving regime, we show that the efficiency can be modulated but will never reach one.
Driven superconducting junctions can still dissipate power due to fluctuations of the superconducting phase, even if the resistance is zero. We therefore also calculate the dissipated power of the driven diode, deriving analytic expressions for the slow and fast driving limits. We show that the dissipation in the ideal diode regime can be several orders of magnitude smaller than in the junction in the normal state for the same dc current and normal resistance.

We exemplify our insights with a simple diode geometry based on a SQUID. Higher harmonics in the current--phase relationship (CPR) can lead to an asymmetric critical current when a magnetic flux penetrates the loop~\cite{Souto_PRL2022}. Several mechanisms can contribute to higher harmonics, including high-transmission modes~\cite{Beenakker_PRL1991} and inductance effects~\cite{Barone_book,Semenov_IEE2015}, studied before in the context of quantum ratchets~\cite{Zapata_PRL1996}. Here, we focus on the latter, which can be experimentally realized using thin, disordered, or granular superconductors~\cite{Rotzinger_SST2017,Niepce_PRAp2019,Moshe_APL2020}. We show that the system can be tuned to the ideal diode regime, with a dissipation that is several orders smaller than in the metallic regime.

The circuit we consider is sketched in Fig.~\ref{fig1}(a) and consists of a superconducting junction (left) that is shunted by a resistor with resistance $R$ (right).
The circuit is connected to an external current source that can be used to bias the circuit simultaneously with an ac and dc current, i.e., $I_{\rm bias}(t) = I_{\rm dc} + I_{\rm ac} \cos(\omega t)$.
Assuming the resistance $R$ to be small enough that the junction is in the overdamped limit, the equation of motion for the superconducting phase difference over the junction $\varphi$ becomes
\begin{equation}
    \dot\varphi + I(\varphi) = I_{\rm dc} + I_{\rm ac} \cos(\omega t),
    \label{Eq:RSJ}
\end{equation}
where all currents have been renormalized by $\hbar/2eR$, which gives them units of $s^{-1}$.
The CPR $I(\varphi)$ of the superconducting junction has to be $2\pi$-periodic and can thus be expanded as
\begin{equation}
    I(\varphi) = \sum_{m \geq 1} I_m \sin(m\varphi + \gamma_m ).\label{eq:cpr}
\end{equation}

In order to have a finite superconducting diode effect at zero driving, i.e., $I_{\rm ac}=0$, the junction must have an asymmetric CPR in such a way that the system has a different maximal current in both directions,
\begin{equation}
    |I_c^+| = \left| \max_\varphi I(\varphi) \right| \neq |I_c^-| = \left| \min_\varphi I(\varphi) \right|.
\end{equation}
In this work, we will not speculate under what physical conditions such a diode effect can arise, but we will treat the set of Fourier coefficients $\{I_m,\gamma_m\}$ as free parameters that can describe any superconducting junction being part of the circuit shown in Fig.~\ref{fig1}(a).
A simple example of a set of coefficients that gives rise to a diode effect is $I_1 = I_2\equiv i_0$, $\gamma_2 = -\frac{\pi}{2}$, and all other coefficients set equal to zero.
This CPR yields $I_c^+ = 2i_0$ and $I_c^- = - \frac{9}{8}i_0$ without current driving, and we will use it below as ``toy'' example to illustrate the response of a superconducting diode to finite ac current driving.

We can analyze the response of the circuit to driving by solving Eq.~\eqref{Eq:RSJ} numerically, see for example Fig.~\ref{fig1}(b), where we show the average differential resistance across the system, $d\langle V\rangle/dI_{\rm dc}$, as a function of the current biases $I_{\rm ac}$ and $I_{\rm dc}$, using $\omega = 0.1i_0$, where $\langle V\rangle$ is the averaged voltage drop over many periods after the steady state is reached.
This figure illustrates one of the
main points we explore in this work:
the zero-voltage superconducting window is tunable via the ac current bias $I_{\rm ac}$.
Moreover, we note that there exists a regime where all zero-voltage supercurrent is positive, and there is one ac bias in particular where the maximum dc current supported at zero voltage is $\tilde I_{\rm dc}^+ > 0$ whereas the minimum dc current is $\tilde I_{\rm dc}^- = 0$, where the tilde indicates that this is the critical dc current under driving. Here, one could say that the system behaves effectively as an ideal diode.
In Fig.~\ref{fig1}(c) we illustrate this regime by plotting the calculated voltage over the junction $\langle V\rangle$ as a function of $I_{\rm dc}$ close to this special point (red trace).
A zero-voltage plateau extends from $I_{\rm dc} \approx 0$ to $I_{\rm dc} \approx 0.8i_0$.
The blue trace shows the $VI$ characteristic at $I_{\rm ac} = 0$, for comparison, confirming the limiting values given by the CPR, $I_c^+ = 2i_0$ and $I_c^- = - \frac{9}{8}i_0$, corresponding to a diode efficiency of $\eta = 0.28$.
Fig.~\ref{fig1}(d) shows the effective efficiency $\tilde \eta=|(\tilde I_{\rm dc}^++\tilde I_{\rm dc}^-)/(\tilde I_{\rm dc}^+-\tilde I_{\rm dc}^-)|$ as a function of $I_{\rm ac}$, as extracted from the data presented in Fig.~\ref{fig1}(b), showing an increase from $\tilde{\eta}=0.28$ to $\tilde \eta=1$, as expected.

Below we will (i) present the simple picture that explains the behavior of the junction in the slow-driving limit $\omega \ll |I_c^\pm|$, explored in Fig.~\ref{fig1}, and (ii) derive analytic understanding of the opposite limit of fast driving, $\omega \gg |I_c^\pm|$, where the extent of the zero-voltage plateau is also tunable via $I_{\rm ac}$, although to a lesser extent.

In the limit of $\omega \ll |I_c^\pm|$, it is helpful to use the common interpretation of Eq.~(\ref{Eq:RSJ}) in terms of an equation of motion of a massless particle with coordinate $\varphi$ in a time-dependent tilted ``washboard'' potential.
We thus write Eq.~(\ref{Eq:RSJ}) in the form $\dot \varphi + \partial_\varphi U(\varphi,t) =0$, where
\begin{equation}
    U(\varphi,t) = -\!\!\sum_{m \geq 1} \frac{I_m}{m} \cos(m \varphi +\gamma_m) - [I_{\rm dc}+ I_{\rm ac} \cos(\omega t)] \varphi.
    \label{eq:potential}
\end{equation}
Since there is no inertia, the particle always adjusts its velocity $\dot\varphi$ instantaneously to the local gradient of the potential, such that the friction force $-\dot\varphi$ and the force $-\partial_\varphi U$ cancel.
A finite average voltage $\langle V\rangle = (\hbar/2e)\langle \dot\varphi \rangle$ corresponds to a finite average velocity of the particle, and in the slow-driving limit $\langle \dot \varphi \rangle \neq 0$ can only arise when there are parts of the driving period where $U$ has no local minima, i.e., it increases or decreases monotonically.
The particle will thus be stuck in the same minimum, resulting in zero average voltage, when the equation $\partial_\varphi U(\varphi,t) = I(\varphi) - I_{\rm dc} - I_{\rm ac} \cos(\omega t) = 0$ has solutions for all $t$.
This leads to the simple condition
\begin{equation}
    I_c^- + I_{\rm ac} < I_{\rm dc} < I_c^+ - I_{\rm ac},
\end{equation}
for the lowest zero-voltage plateau; the boundaries of this region are indicated by the red dashed lines in Fig.~\ref{fig1}(b).
The effective diode efficiency at finite $I_{\rm ac}$ follows as
\begin{equation}
    \tilde \eta = \frac{I_c^+ + I_c^-}{I_c^+ - I_c^- - 2 I_{\rm ac}},
    \label{Eq:eta_driving}
\end{equation}
from which we find that the maximal efficiency $\tilde \eta = \pm 1$ occurs simply when $I_{\rm ac} = \min \{ |I_c^-|, |I_c^+| \}$.
For larger $I_{\rm ac}$, the zero-resistance window does not contain $I_{\rm dc}=0$ and the interpretation of $\tilde \eta$ as an effective diode efficiency becomes meaningless, gray area in Fig.~\ref{fig1}(d).

\begin{figure}[t]
\begin{center}
    \includegraphics[width=1.0\linewidth]{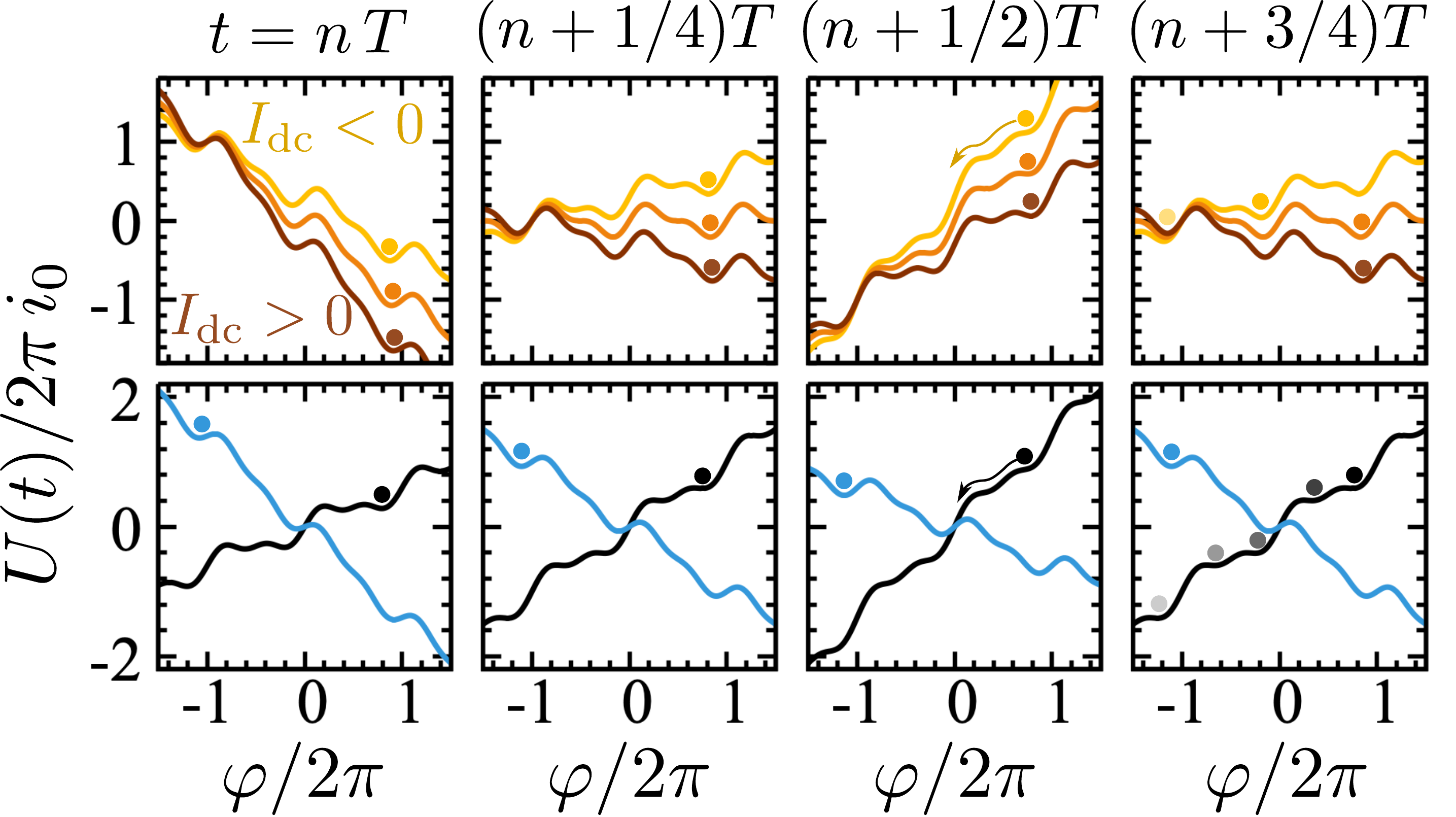}
\end{center}
\caption{{\bf Rocking washboard potential.} Effective time-dependent potential describing the Josephson diode under an ac current drive, $I=I_{\rm dc}+I_{\rm ac}\cos(\omega t)$, where the dots illustrate the phase dynamics. Upper panels show the situation where the diode is in the ideal regime with $I_{\rm dc}=-0.3i_0$ (yellow), $0$ (orange), and $0.3i_0$ (brown). The phase stays in the same potential minimum after a cycle for $I_{\rm dc}\geq0$, while it can drift for $I_{\rm dc}<0$, ending up in a different minimum after each period. Lower panels correspond to the black and blue dots in Fig.~\ref{fig1}(b), $I_{\rm ac}=0.4i_0$ and $I_{\rm dc}=\pm i_0$. For $I_{\rm dc}=i_0$ (blue) the potential always has local minima and the phase is trapped. In contrast, the phase can drift for $I_{\rm dc}=-i_0$. Depending on the driving frequency, the phase can stay or drift between neighboring minima (black and gray dots), leading to the doubling of the Shapiro steps shown in Fig.~\ref{fig1}(b).
\label{fig:potentials}}
\end{figure}

Let us illustrate this picture for the example CPR considered in Fig.~\ref{fig1}, where the effective efficiency reaches $\tilde \eta = 1$ for $I_{\rm ac}= |I_c^-|$.
For this particular driving strength, the behavior of the time-dependent potential $U(\varphi,t)$ is sketched in the top row of Fig~\ref{fig:potentials} for $I_{\rm dc}>0$, $I_{\rm dc}=0$, and $I_{\rm dc}<0$ (brown, orange, and yellow curves, respectively).
The driving provides a time-dependent tilt of the potential; the potentials with maximal negative ($t=n\,T$) and positive [$t=(n+\frac{1}{2})T$] tilt per period are shown in the first and third columns of Fig.~\ref{fig:potentials}.
For an adiabatic drive, the phase would be able to adapt to the instantaneous potential,
and a finite average voltage can thus develop as soon as the potential becomes monotonous for part of the driving period.
For $I_{\rm dc} = 0$ (orange curve), this happens exactly at $t=(n+\frac{1}{2})T$, where the tilt is maximal.
A small additional current bias $I_{\rm dc}\neq 0$ yields an extra time-independent tilt $-I_{\rm dc}\varphi$ in the potential.
If $I_{\rm dc} > 0$ (brown curve), the extra dc tilt 
restores the barriers at $t=(n+\frac{1}{2})T$ resulting in the particle being trapped in the same minimum during the whole period
and $\langle \dot \varphi \rangle=0$. In contrast, any $I_{\rm dc} < 0$ (yellow curve) would increase the slope of the curve resulting in a part of the driving period around $t=(n+\frac{1}{2})T$ where the potential has no local minima and maxima, allowing a finite average velocity $\langle \dot \varphi \rangle < 0$ which makes the system resistive. This illustrates the physics of the $\tilde{\eta}=1$ case, where a finite voltage drop appears only for one direction of the dc current.

In reality, driving is never truly in the (infinitely) slow limit, and most additional structure seen in Fig.~\ref{fig1}(b) can be attributed to finite-frequency effects.
The Shapiro steps observed outside the zero-voltage region reflect the fact that the number of local minima that the phase can drift per period is discrete, as illustrated by the dots in the right panels of Fig.~\ref{fig:potentials}, and this number increases with increasing $I_{\rm ac}$.
The apparent doubling of the number of Shapiro steps at $I_{\rm dc} < -I_{\rm ac}$ is a result of the specific shape of our CPR~\cite{Zapata_PRL1996}:
The time-dependent potential for a point in the region with additional Shapiro steps [black dot in Fig.~\ref{fig1}(b)] is illustrated by the black curve in the lower panels of Fig.~\ref{fig:potentials}.
In this case, the black potential with the smallest average slope has \emph{two} inequivalent local minima.
Considering the driving frequency to be finite, the first dissipative processes yield a phase jump between the different potential minima (gray dots).
It then takes two periods to change the phase by an integer of $2\pi$, whereas in Fig.~\ref{fig:potentials}(a) the first available process changes the phase by a multiple of $2\pi$ each period.
This explains the doubling of the Shapiro steps shown in Fig.~\ref{fig1}(b). For comparison, we show in blue the situation with the opposite $I_{\rm dc}$ [blue dot in Fig.~\ref{fig1}(b)], where the phase does not drift, implying an asymmetry of the Josephson potential. All this means that the detailed structure of the Shapiro steps in the $(I_{\rm ac},I_{\rm dc})$-plane encodes information about the shape of the CPR.

\begin{figure}[t!]
\begin{center}
    \includegraphics[width=1.0\linewidth]{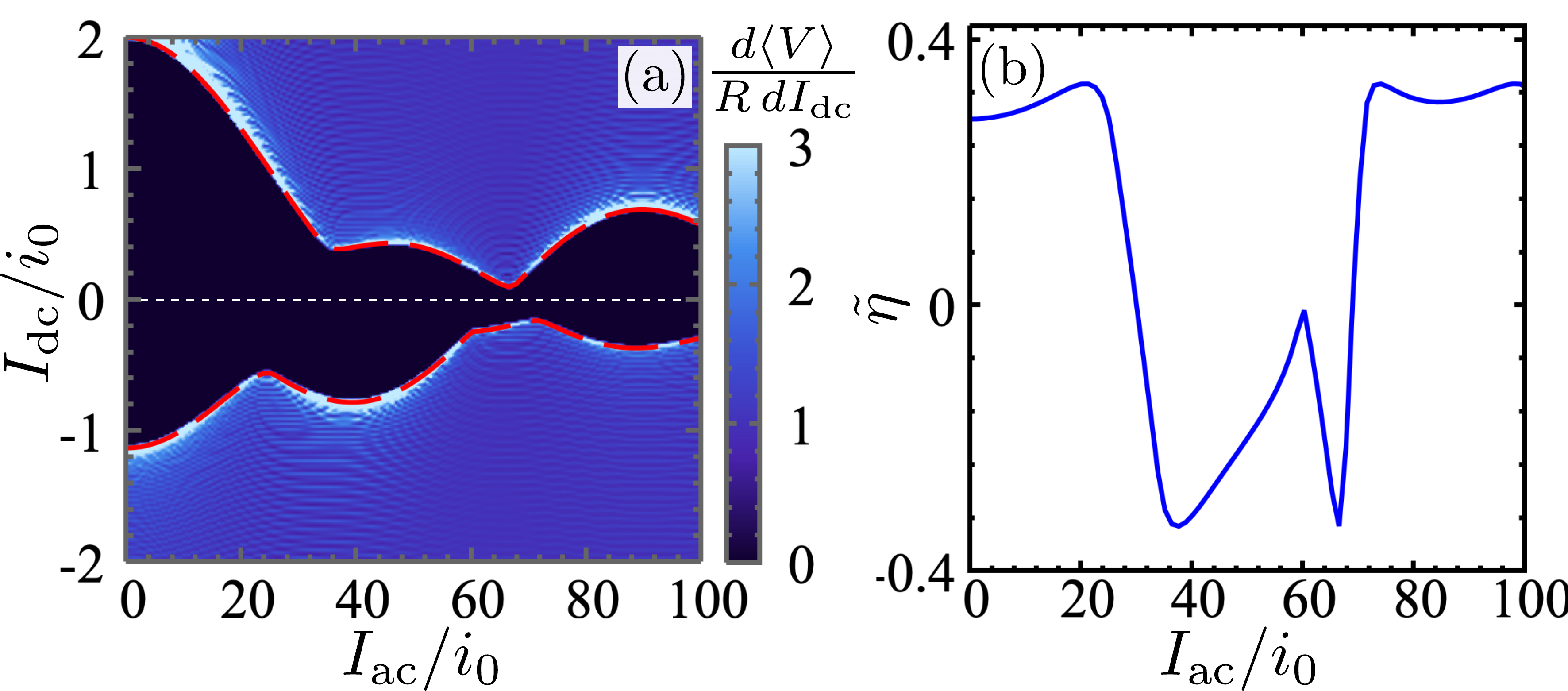}
\end{center}
\caption{{\bf Diode response in the fast-driving regime.} (a) Differential resistance as a function of the dc and ac bias currents. The red dashed lines correspond to the critical currents predicted by Eq.~(\ref{eq:drivenIdc}). (b) Effective diode efficiency as a function of $I_{\rm ac}$. In both panels we use the same CPR as in Fig.~\ref{fig1} and we set $\omega=25i_0$.
\label{fig2}}
\end{figure}

In the case of fast driving, $\omega \gg |I_c^\pm|$, the reaction of the phase to the driving is more complex.
Still, the system can show a region with an average zero voltage drop, as shown in Fig.~\ref{fig2}(a), and we can derive analytic expressions describing the boundaries of this region.

In the zero voltage drop situation, the phase is periodic in time and can be written as
\begin{equation}
\varphi(t) = \alpha_0 + \sum_{n\geq 1} \alpha_n \cos(n\omega t + \beta_n).
\label{eq:phifast}
\end{equation}
Inserting this Fourier expansion into the CPR (\ref{eq:cpr}), we find that we can write
\begin{align}
	  I(\varphi) = {} & {} \sum_{m\geq 1} I_m {\rm Im} \Bigg[ e^{i\gamma_m} \prod_{n\geq 0} \sum_p i^p J_p(m\alpha_n) e^{i p n\omega t}e^{i p \beta_n} \Bigg],
\end{align}
with $J_p(x)$ being the $p$-th Bessel function of the first kind and using $\beta_0 = 0$.
We can then inspect Eq.~(\ref{Eq:RSJ}) and equate all zero-frequency terms, which yields the equation $I_{\rm dc} = \sum_{m\geq 1} I_m \sin(m\alpha_0+\gamma_m)\prod_{n\geq 1} J_0(m\alpha_n)$.
Equating all terms in Eq.~(\ref{Eq:RSJ}) that oscillate with frequency $\omega$, and then all terms that oscillate with $n\omega$ where $n>1$, we find that in the limit $\omega \gg I_m$ we have $I_{\rm ac} \approx \omega \alpha_1$ and all $\alpha_n$ with $n>1$ can be neglected.
This finally yields
\begin{equation}
    I_{\rm dc} = \sum_{m\geq 1} I_m \sin(m\alpha_0+\gamma_m) J_0\left(m\frac{I_{\rm ac}}{\omega}\right),
    \label{eq:drivenIdc}
\end{equation}
and the boundaries of the zero-voltage region are the maximal and minimal dc current that can be supported according to this equation by adjusting $\alpha_0$.
We thus see that in the fast-driving limit the leading effect of the driving is a ``dressing'' of all Fourier components of the CPR by a factor $J_0(mI_{\rm ac}/\omega)$.

The red dashed lines in Fig.~\ref{fig2}(a) show the maximal and minimal zero-voltage dc currents $I_{\rm dc}^\pm$ found from Eq.~(\ref{eq:drivenIdc}) for the chosen parameters (see captions of Figs.~\ref{fig1} and \ref{fig2}).
For this particularly simple CPR, analytic expressions for the critical currents can in fact be derived (see the Appendix), but in general one needs to find the extrema of the dressed CPR (\ref{eq:drivenIdc}) numerically.
The corresponding effective diode efficiency as a function of $I_{\rm ac}$ is shown in Fig.~\ref{fig2}(b), which is clearly tunable through $I_{\rm ac}$ but to a lesser extent than in the slow-driving case.
We note that the directionality of the diode behavior can be changed by tuning $I_{\rm ac}$.
The efficiency does not exceed $|\tilde\eta| \approx 0.34$, but we still see that it can be improved by approximately 10\% as compared to the non-driven case.
Indeed, fast driving cannot tune the diode to become ideal ($\tilde \eta=1$)
since also the dressed CPR is zero on average.

\begin{figure}[t]
\begin{center}
    \includegraphics[width=1.0\linewidth]{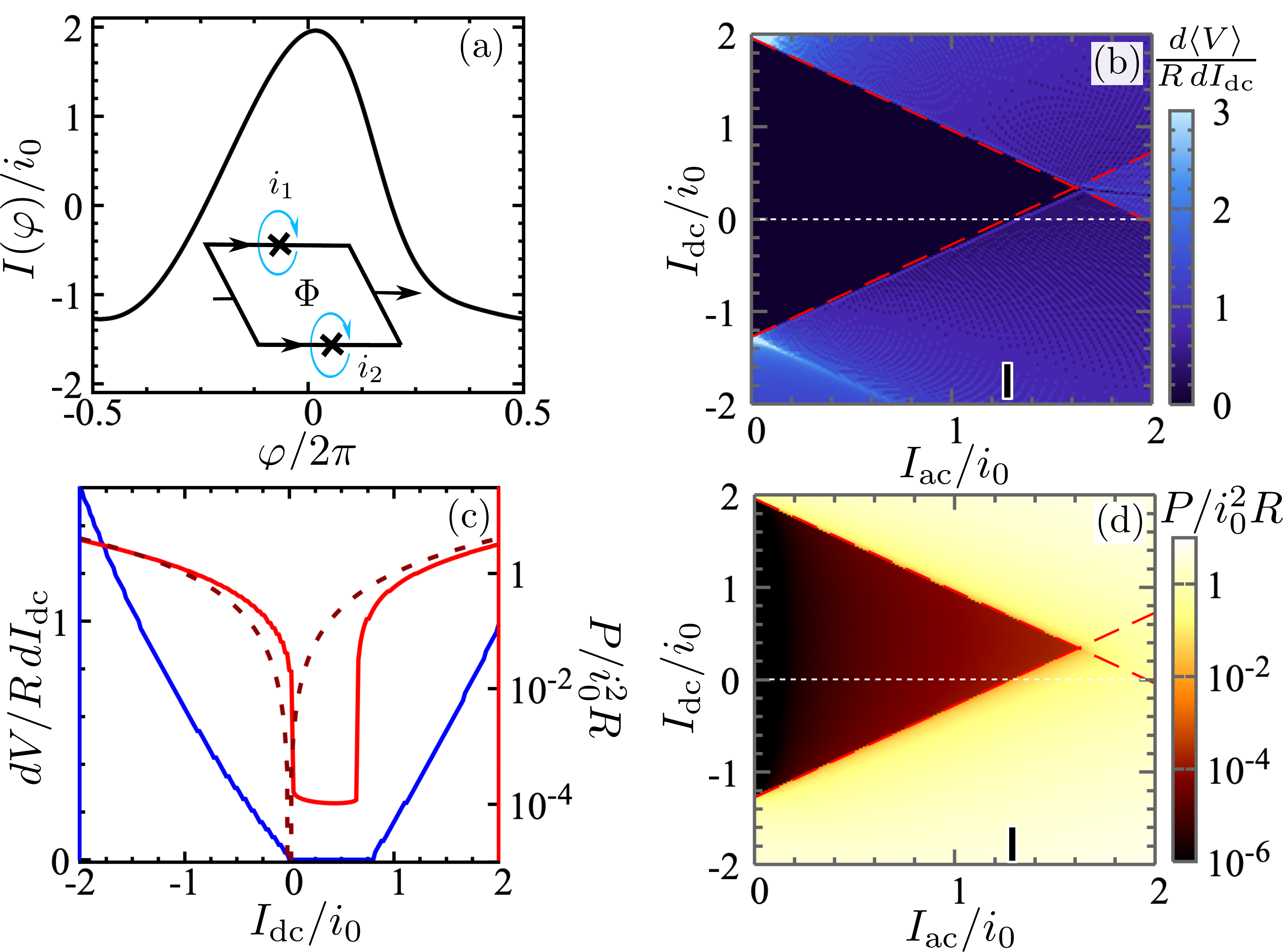}
\end{center}
\caption{{\bf Inductance diode.} (a) CPR. Inset: sketch of the system, consisting of two conventional Josephson junctions in a SQUID geometry, where the arms of the loop have a finite inductance. (b) Differential resistance. (c) Resistance (blue) and dissipated Joule power (solid red) as a function of the dc current for $\tilde{I}_{\rm ac}=1.32i_0$. Dashed line shows the dissipated power in the normal state. (d) Dissipated power as a function of the ac and dc bias current. The critical currents of the two junctions are taken as $i_{1c}=i_{2c}/2 = i_0$, $L=0.2\hbar/e\,i_0$, and the flux due to the external magnetic field $\Phi_E/\Phi_0=0.35$, with $\Phi_0$ being the flux quantum.\label{fig4}}
\end{figure}

In all cases considered, even if the average voltage drop across the junction, and therefore the calculated differential resistance, is zero, the driven diode can still dissipate energy. This occurs because of shifts in the phase $\varphi$ resulting from alterations in the location of the potential minima.
To quantify this effect, we calculate the average dissipated power as 
\begin{equation}
    P=\frac{1}{T}\int_{0}^{T} dt\, V(t) I_{\rm bias}(t),\label{eq:p}
\end{equation}
where we integrate over one period, assuming that the system has reached steady state.
In the slow-driving regime, the residual dissipation is due to 
changes of the effective potential that shifts the minima.
Considering the zero-voltage region, where the phase stays trapped in the same local minimum, we can expand the potential around this minimum as $U(\varphi, t) \approx U_0 + \alpha (\varphi - \varphi_{\rm min})^2 - I_{\rm bias}(t)\varphi$ and derive from Eq.~(\ref{Eq:RSJ}) (see the Appendix)
\begin{equation}
    P_{\rm slow}\approx\frac{\omega^2 I^2_{\rm ac}R}{8\alpha^2}.\label{eq:pslow}
\end{equation}
From this expression, it is clear that decreasing the driving frequency or increasing the sharpness of the confining potential, $\alpha$, will decrease dissipation.
In the fast-driving regime, one can use Eq.~(\ref{eq:phifast}) to find
\begin{equation}
    P_{\rm fast}\approx\frac{I^2_{\rm ac}R}{2}.\label{eq:pfast}
\end{equation}
In Fig.~S5 of the Appendix, we compare dissipation in the slow- and the fast-driving regimes and show that the two approximate expressions (\ref{eq:pslow},\ref{eq:pfast}) fit numerical results well.
We see that the system will dissipate much more energy in the fast-driving than in the slow-driving regime, even in the regime where differential resistance is zero. Indeed, the phase remains in the same potential well (zero voltage drop in average), although it cannot adapt to the minimum instantaneously, leading to a phase amplitude proportional to $I_{\rm ac}$, and a dissipated power $\propto I^2_{\rm ac}$.

Finally, we discuss a physical implementation where our predictions can be tested, although the described mechanism is general and compatible with all diode proposals based on dc mechanisms. Here, we consider a SQUID with two arms that have a finite inductance, as illustrated in Fig.~\ref{fig4}(a), a well understood device~\cite{Barone_book}. In that figure, we use an inductance for each arm that has a value of $L=0.2\hbar/e\,i_0$ that would correspond to $\sim1$nH for a critical current of $\sim 100$nA. This condition can be easily reached in superconductor-semiconductor junction where the critical current is tunable. For simplicity, we consider that the two junctions have a sinusoidal CPR. The inductance leads to the onset of higher harmonics in the CPR of each of the junctions. We find an asymmetry in the SQUID's critical current whenever the critical currents or the inductances of the two junctions are different and there is a finite magnetic flux penetrating the loop. The CPR is shown in Fig.~\ref{fig4}(a), indeed showing an asymmetry between negative and positive directions. The calculated differential resistance under driving is shown in Fig.~\ref{fig4}(b) for the slow-driving regime, showing signatures that are qualitative similar to those of the simple diode example of Fig.~\ref{fig1}(c). The blue curve in Fig.~\ref{fig4}(c) shows a line cut of the differential resistance at $I_{\rm ac} = 1.32i_0$, where $\tilde{\eta}=1$ for a realistic diode implementation.
In Fig.~\ref{fig4}(d) we show the dissipated power for the inductance-based diode in the slow-driving regime, as calculated from Eq.~(\ref{eq:p}), using a logarithmic color scale.
The dissipated power is strongly suppressed in the zero differential resistance region, see Fig.~\ref{fig4}(c), where it is approximately given by Eq.~(\ref{eq:pslow}).
In Fig.~\ref{fig4}(c) we also show a line cut of the dissipated power at $\tilde \eta = 1$ (solid red line). For illustration, we show the dissipated power by the device in the normal state with $I_{\rm ac}=0$ (dashed line). We note a reduction of several orders of magnitude, demonstrating that ideal diodes are promising for low-dissipation circuit elements.

To conclude, in this article we studied the possibility to control the superconducting diode response using ac current driving. We have shown that a slow drive can tune the system to a situation where the zero-resistance plateau extends in only one direction, yielding an effective efficiency of 1, independent of the diode origin. We illustrated this possibility in a realistic implementation of a SQUID diode based on Josephson junctions with significant inductance. Another example can be found in Ref.~\cite{Valentinin_arXiv2023}, where the SQUID Josephson junctions themselves feature higher harmonics. In the opposite regime of high-frequency driving, the drive can still tune the diode response, but without ever reaching efficiency 1.
In all cases, the system will dissipate energy under driving due to phase fluctuations in time. We have shown that in the slow-driving regime this dissipation can be small, opening the door for new low-dissipation circuit elements. The proposal and realization of an ideal superconducting diode is an open challenge in the field. 

{\it Acknowledgements.-} We acknowledge support from research grants Spanish CM “Talento Program” (project No. 2022-T1/IND-24070), Spanish Ministry of Science, innovation, and Universities through Grant PID2022-140552NA-I00, Swedish Research Council under Grant Agreement No. 2020-03412, the European Research Council (ERC) under the European Union’s Horizon 2020 research and innovation programme under Grant Agreement No. 856526., Nanolund, FWF Project F-8606, and Microsoft Corporation.

\appendix

\begin{widetext}

\section{Boundaries of the zero-voltage region under fast driving}

The differential equation to solve is
\begin{equation}
    \dot\varphi + I(\varphi) = I_{\rm dc} + I_{\rm ac} \cos(\omega t),\label{eq:diffeq}
\end{equation}
with all currents renormalized by $\hbar/2eR$.
We assume the phase to be periodic in steady state, 
\begin{equation}
    \varphi(t) = \alpha_0 + \sum_{n\geq 1} \alpha_n \cos(n\omega t + \beta_n).
\end{equation}
which yields
\begin{align}
	  \dot\varphi = {} & {} -\sum_{n\geq 1} n \alpha_n \omega \sin(n\omega t + \beta_n),\label{eq:phidot}
\end{align}
as well as
\begin{align}
	  I(\varphi) = {} & {} \sum_{m \geq 1} I_m \cos\left[m\varphi + \gamma_m \right]\nonumber\\
	  = {} & {} \sum_{m\geq 1} \frac{I_m}{2} \Bigg[ e^{i(m\alpha_0 + \gamma_m)} \prod_{n\geq 1} \sum_p i^p J_p(m\alpha_n) e^{i p n\omega t}e^{i p \beta_n} + e^{-i(m\alpha_0 + \gamma_m)} \prod_{n\geq 1} \sum_p i^p J_p(-m\alpha_n) e^{i p n\omega t}e^{i p \beta_n} \Bigg].
\end{align}
This finally allows us to rewrite Eq.~(\ref{eq:diffeq}) as
\begin{align}
	  I_{\rm dc} + \frac{I_{\rm ac}}{2}\left( e^{i\omega t} + e^{-i\omega t} \right) = {} & {} 
	  \frac{i\omega}{2} \sum_{n\geq 1}n\alpha_n \left( e^{in\omega t}e^{i\beta_n} - e^{-in\omega t}e^{-i\beta_n} \right) \nonumber\\
	  {} & {} + \sum_{m\geq 1} \frac{I_m}{2} \Bigg[ e^{i(m\alpha_0 + \gamma_m)} \prod_{n\geq 1} \sum_p i^p J_p(m\alpha_n) e^{i p n\omega t}e^{i p \beta_n}
   \nonumber \\ {} & {} \hspace{5em} 
   + e^{-i(m\alpha_0 + \gamma_m)} \prod_{n\geq 1} \sum_p i^p J_p(-m\alpha_n) e^{i p n\omega t}e^{i p \beta_n} \Bigg].\label{eq:full}
\end{align}
  
We then start equating the Fourier components on the left and right hand side of this equation.
The constant part of the equation ($\propto 1$) becomes
\begin{align}
			  I_{\rm dc}  =  \sum_{m\geq 1} I_m \cos(m\alpha_0 + \gamma_m) \prod_{n\geq 1}  J_0(m\alpha_n),
\end{align}
which presents the modified effective current--phase relationship under driving.
The maximum and minimum dc currents follow from maximizing this expression over all free parameters $\alpha_n$.

We then consider the first Fourier component of the equation ($\propto e^{i\omega t}$), yielding
\begin{align}
			  \frac{I_{\rm ac}}{2}= {} & {} 
			  \frac{i\omega}{2} \alpha_1 e^{i\beta_1} + \sum_{m\geq 1} \frac{I_m}{2} \Bigg[ e^{i(m\alpha_0 + \gamma_m)} \prod_{r\geq 1} \sum_p i^p J_p(m\alpha_r) e^{i p r\omega t}e^{i p \beta_r} 
     \nonumber\\  {} & {} \hspace{9em} 
     + e^{-i(m\alpha_0 + \gamma_m)} \prod_{r\geq 1} \sum_p i^p J_p(-m\alpha_r) e^{i p r\omega t}e^{i p \beta_r} \Bigg]_{\propto e^{i\omega t}},
\end{align}
where the final subscript means that only the terms proportional to $e^{i\omega t}$ should be considered.
In the limit $\omega \gg I_m$ we can reduce this equation to an equation for $\alpha_1$ only,
\begin{equation}
    I_{\rm ac} \approx i\omega \alpha_1 e^{i\beta_1}.
\end{equation}
Working with $I_{\rm ac},\omega >0$, we see that $\beta = -\pi/2$ makes $\alpha_1>0$, such that we can write $\alpha_1 \approx I_{\rm ac}/\omega$.
The terms proportional to $e^{-i\omega t}$ in Eq.~(\ref{eq:full}) give the same result and all higher Fourier components in the equation $\propto e^{i n\omega t}$ with $|n| > 1$ only produce small corrections to the $\alpha_n$ with $n>1$.

We thus approximate $\alpha_1 = I_{\rm ac}/\omega$ and $\beta_1 = -\pi/2$ and find
\begin{align}
	  I_{\rm dc}  \approx {} & {} \sum_{m\geq 1} I_m \cos(m\alpha_0 + \gamma_m) J_0\left(m\frac{I_{\rm ac}}{\omega}\right),\label{eq:idc}
\end{align}
where $\alpha_0$ is the only remaining free parameter, over which $I_{\rm dc}$ can be maximized and minimized to find the boundaries of the zero-voltage region,
\begin{align}
	  I_{\rm dc}^+ = {} & {} {\rm max}_{\alpha_0} \left\{ \sum_{m\geq 1} I_m \cos(m\alpha_0 + \gamma_m) J_0\left(m\frac{I_{\rm ac}}{\omega}\right)\right\},\label{eq:icplus}\\
	  I_{\rm dc}^- = {} & {} {\rm min}_{\alpha_0} \left\{ \sum_{m\geq 1} I_m \cos(m\alpha_0 + \gamma_m) J_0\left(m\frac{I_{\rm ac}}{\omega}\right)\right\}.\label{eq:icminus}
\end{align}

For a simple current--phase relationship, such as
\begin{equation}
    I(\varphi) =i_0\sin(\varphi)+i_0\sin(2\varphi-\tfrac{\pi}{2}),
\end{equation}
as we used as example in the main text, one can find analytic expressions for the boundaries of the zero-voltage,
\begin{align}
    I_{\rm dc}^+ = {} & {}
    \begin{cases}
    {\displaystyle -\frac{J_0(i_{\rm ac})^2}{8J_0(2i_{\rm ac})} - J_0(2i_{\rm ac})} & \text{for $|J_0(i_{\rm ac})| < -4 J_0(2i_{\rm ac})$},\\[.8em]
    {\rm max} \big\{ J_0(2i_{\rm ac}) \pm J_0(i_{\rm ac}) \big\} & \text{elsewhere},
    \end{cases}\\
    I_{\rm dc}^- = {} & {}
    \begin{cases}
    {\displaystyle -\frac{J_0(i_{\rm ac})^2}{8J_0(2i_{\rm ac})} - J_0(2i_{\rm ac})} & \text{for $|J_0(i_{\rm ac})| < 4 J_0(2i_{\rm ac})$},\\[.8em]
    {\rm min} \big\{ J_0(2i_{\rm ac}) \pm J_0(i_{\rm ac}) \big\} & \text{elsewhere},
    \end{cases}
\end{align}
where we used $i_{\rm ac} = I_{\rm ac}/\omega$.

\section{Calculation of the steady-state dissipation}

In this Section, we will explain how we arrived at the two approximate expressions for the average power dissipated in the fast- and slow-driving limits.
In all cases, we calculate the dissipation as
\begin{align}
     P = {} & {} \frac{1}{T}\int_{0}^{T} dt\, V(t) I_{\rm bias}(t)
     = \frac{1}{T}\int_{0}^{T} dt\, R \dot \varphi [I_{\rm dc}+ I_{\rm ac} \cos(\omega t)],\label{eq:pint}
\end{align}
where we used that $V(t) = R\dot\varphi$ since $\dot\varphi$ was defined to be renormalized by $\hbar/2eR$.

\subsection{Fast-driving regime}

In the fast-driving limit, we can use the results from the previous section.
We see from Eq.~(\ref{eq:phidot}) that
\begin{equation}
    \dot \varphi \approx -I_{\rm ac} \sin(\omega t - \tfrac{\pi}{2} ),
\end{equation}
where we made use of the result $\alpha_1 = I_{\rm ac}/\omega$ and $\beta_1 = -\pi/2$.
We then straightforwardly find from Eq.~(\ref{eq:pint})
\begin{align}
     P  = \frac{I_{\rm ac}^2R}{2}.\label{eq:pfast}
\end{align}

\subsection{Slow-driving regime}

In this case, we will assume that the driving is slow enough such that the phase will more or less constantly follow the minimum of the time-dependent potential $U(\varphi,t)$.
The equation of motion reads as
\begin{equation}
    \dot \varphi =- \partial_\varphi U(\varphi,t),
\end{equation}
with
\begin{equation}
     U(\varphi,t) = - \sum_{m \geq 1} \frac{I_m}{m} \cos(m \varphi +\gamma_m) - [I_{\rm dc}+ I_{\rm ac} \cos(\omega t)] \varphi.
    \label{eq:potential}
\end{equation}
The first part of the potential (i.e., the effective Josephson energy of the whole junction) can be expanded around its minimum,
\begin{equation}
     U(\varphi,t) \approx U_0 + \alpha (\varphi - \varphi_0)^2 - [I_{\rm dc}+ I_{\rm ac} \cos(\omega t)] \varphi.\label{eq:potapp}
\end{equation}

As a zeroth approximation, we will assume that the phase follows the minimum exactly and instantaneously (the minimum will be periodically shifted back and forth by the rest of the rocking potential).
We find the time-dependent minimum of the potential from solving $\partial_\varphi U(\varphi,t) = 0$, yielding
\begin{equation}
    \varphi_{\rm min}(t) = \varphi_0 + \frac{I_{\rm dc}+ I_{\rm ac} \cos(\omega t)}{2\alpha},
\end{equation}
and we thus assume that
\begin{equation}
    \dot\varphi(t) = \dot \varphi_{\rm min}(t) = -\frac{I_{\rm ac} \omega}{2\alpha} \sin(\omega t).
\end{equation}
Inserting this approximation in Eq.~(\ref{eq:pint}), we see that it yields $P=0$.
Indeed, by insisting that the phase always exactly follows the potential minimum, it never needs to ``slide down'' the potential, which is the process that yields dissipation.
The ``particle'' trapped in the rocking potential will only periodically absorb and release energy as its potential energy is raised and lowered by the rocking.

In order to describe dissipation we thus need to acknowledge that as soon as $\omega > 0$ the phase will always be slightly behind in following the potential minimum, and is thus effectively constantly sliding down the potential while trying to reach the instantaneous minimum.
To describe this effect, we first consider a potential with an initial minimum at $\varphi_0$ and a constant increase of the tilt of the potential with rate $\eta$,
\begin{equation}
    U(\varphi,t) = U_0 + \alpha(\varphi - \varphi_0)^2 + \eta t \varphi.\label{eq:potentialrate}
\end{equation}
The equation of motion now becomes
\begin{align}
    \dot\varphi = {} & {} -2\alpha (\varphi - \varphi_0) + \eta t.
\end{align}
which can be solved, yielding
\begin{equation}
    \varphi(t) = \varphi_0 -\frac{\eta t}{2\alpha} + \frac{\eta}{4\alpha^2} ( 1 - e^{-2\alpha t} ),
\end{equation}
where we used that $\varphi(0) = \varphi_0$.
At long times, $t \gg 1/\alpha$, we thus find
\begin{equation}
    \varphi(t) \approx \varphi_0 -\frac{\eta t}{2\alpha} + \frac{\eta}{4\alpha^2}.\label{eq:phit}
\end{equation}
From Eq.~(\ref{eq:potentialrate}) it is easy to find that the instantaneous position of the minimum of the potential is $\varphi_{\rm min}(t) = \varphi_0 - \eta t / 2\alpha$ in this case.
Comparing this with Eq.~(\ref{eq:phit}), we find
\begin{equation}
    \varphi(t) \approx \varphi_{\rm min}(t) + \frac{\eta}{4\alpha^2},
\end{equation}
i.e., the phase will follow the minimum of the potential at a distance given by $\eta / 4\alpha^2$.

Now we again focus on the rocking potential given in Eq.~(\ref{eq:potapp}). 
We assume that the modulation of the potential is slow enough such that we are always effectively in the long-time limit where the phase has its dynamics adjusted to the slowly changing instantaneous tilting speed $\eta (t)$.
We thus identify
\begin{equation}
    \eta (t) = \omega I_{\rm ac} \sin(\omega t),
\end{equation}
yielding
\begin{equation}
    \varphi(t) \approx \varphi_{\rm min}(t) + \frac{\omega I_{\rm ac}}{4\alpha^2}\sin(\omega t)
    = \varphi_0 + \frac{I_{\rm dc}+ I_{\rm ac} \cos(\omega t)}{2\alpha} + \frac{\omega I_{\rm ac}}{4\alpha^2}\sin(\omega t),
\end{equation}
where we also substituted for $\varphi_{\rm min}(t)$ the minimum of the rocking potential.
This allows for a straightforward calculation of the dissipation,
\begin{align}
     P= \frac{\omega^2 I^2_{\rm ac}R}{8\alpha^2}.\label{eq:pslow}
\end{align}

\section{Current--phase relationship for a SQUID with significant inductance}

\begin{figure}[t]
\begin{center}
    \includegraphics[width=0.2\linewidth]{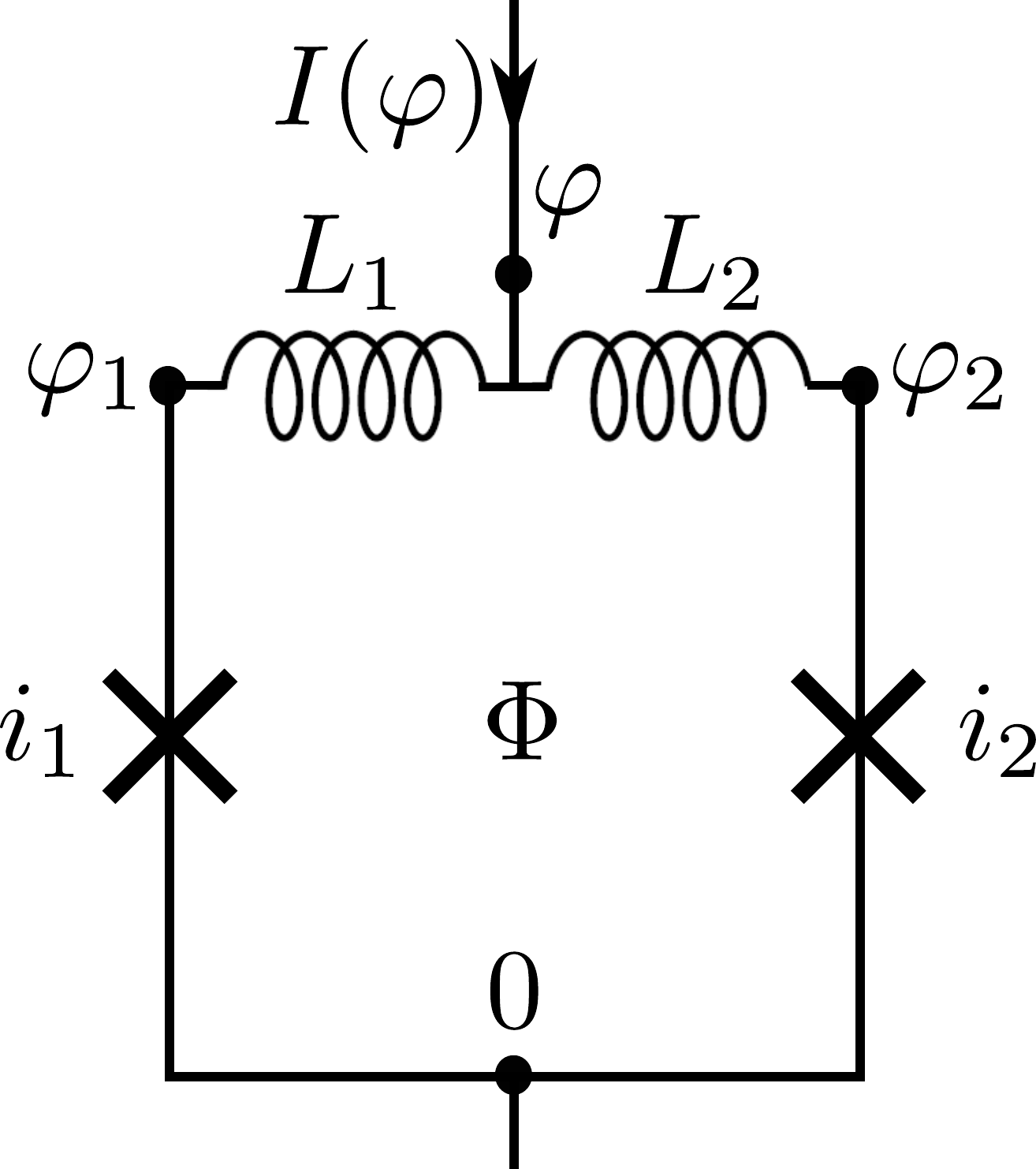}
\end{center}
\caption{Schematic of the model circuit we use to derive the current--phase relationship of the inductance diode.
\label{fig:circuitIncuctance}}
\end{figure}

We consider the circuit model of a SQUID shown in Fig.~\ref{fig:circuitIncuctance}, where both arms of the SQUID have a significant inductance, denoted by $L_{1,2}$ and separated from the junction for simplicity.
The total current can be written as the sum over the two currents $i_{1,2}$ through the Josephson junctions,
\begin{equation}
    I(\varphi)=i_1(\varphi_1)+i_2(\varphi_2),\label{eq:iphi}
\end{equation}
where $\varphi_j$ is the phase difference over the given junction, as indicated in the Figure.
The phases can be related as
\begin{eqnarray}
    \varphi&=&\varphi_1+L_1 i_1(\varphi_1),\\
    \varphi&=&\varphi_2+L_2 i_2(\varphi_2)-2\pi \frac{\Phi}{\Phi_0},
\end{eqnarray}
where we have included the effect of the finite flux penetrating the loop into the phase difference over junction $2$.
For simplicity, we consider junctions with purely sinusoidal current--phase relationships, given by 
\begin{equation}
    i_j(\varphi_j)=i^{c}_j \sin(\varphi_j).\label{eq:ij}
\end{equation}
Eqs.~(\ref{eq:iphi}--\ref{eq:ij}) together implicitly define the current--phase relationship $I(\varphi)$ of the whole circuit.

\section{Supporting results}

\begin{figure}[t]
\begin{center}
    \includegraphics[width=1\linewidth]{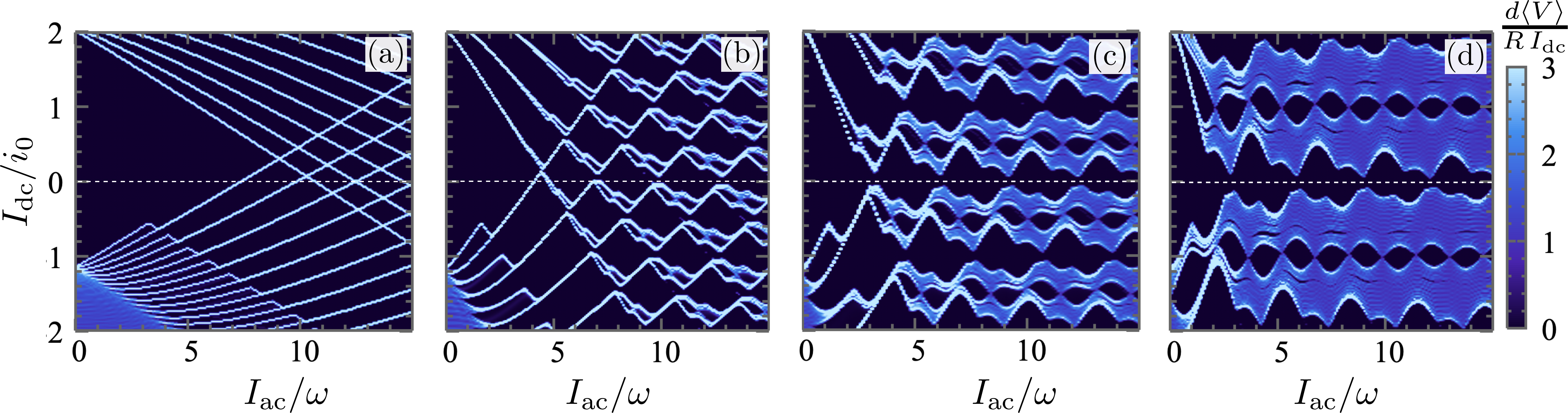}
\end{center}
\caption{From adiabatic to fast driving. Panels (a--d) correspond to $\omega/i_0=0.2$, $0.5$, $1$, and $2$, respectively. We use the same current--phase relationship as in Fig.~1 of the main text, $I(\varphi)=i_0\sin(\varphi)+i_0\sin(2\varphi-\frac{\pi}{2})$.
\label{Fig:adiabaticToFast}}
\end{figure}

In this Section, we present results that complement the ones presented in the main text.
In Fig.~\ref{Fig:adiabaticToFast} we show the differential resistance, numerically calculated as the derivative of the averaged voltage drop with respect to the dc current, for the driving frequencies $\omega/i_0=0.2$, $0.5$, $1$, and $2$, which span the crossover from slow to fast driving.
In the slow-driving regime, the differential resistance shows sharp steps, related to the phase dynamics when the phase can overcome a certain number of potential barriers during each driving period, see Fig.~\ref{Fig:adiabaticToFast}(a) (and Fig.~1 in the main text).
In the intermediate regime, the steps acquire additional structure, becoming progressively broader when increasing the driving frequency $\omega$, see Figs.~\ref{Fig:adiabaticToFast}(b,c).
Finally, for $\omega\gtrsim i_0$ we approach the fast-driving limit described above, the properties of the dc supercurrent then being captured approximately by Eq.~(\ref{eq:idc}). 
In this limit, the critical currents in the two directions oscillate as a function of the driving amplitude $I_{\rm ac}$, see Fig.~\ref{longRenageFast}(a). We see that, even though the critical current can be modulated using the ac current, the efficiency cannot be boosted significantly, as illustrated in Fig.~\ref{longRenageFast}(b).

\begin{figure}[h]
\begin{center}
    \includegraphics[width=.5\linewidth]{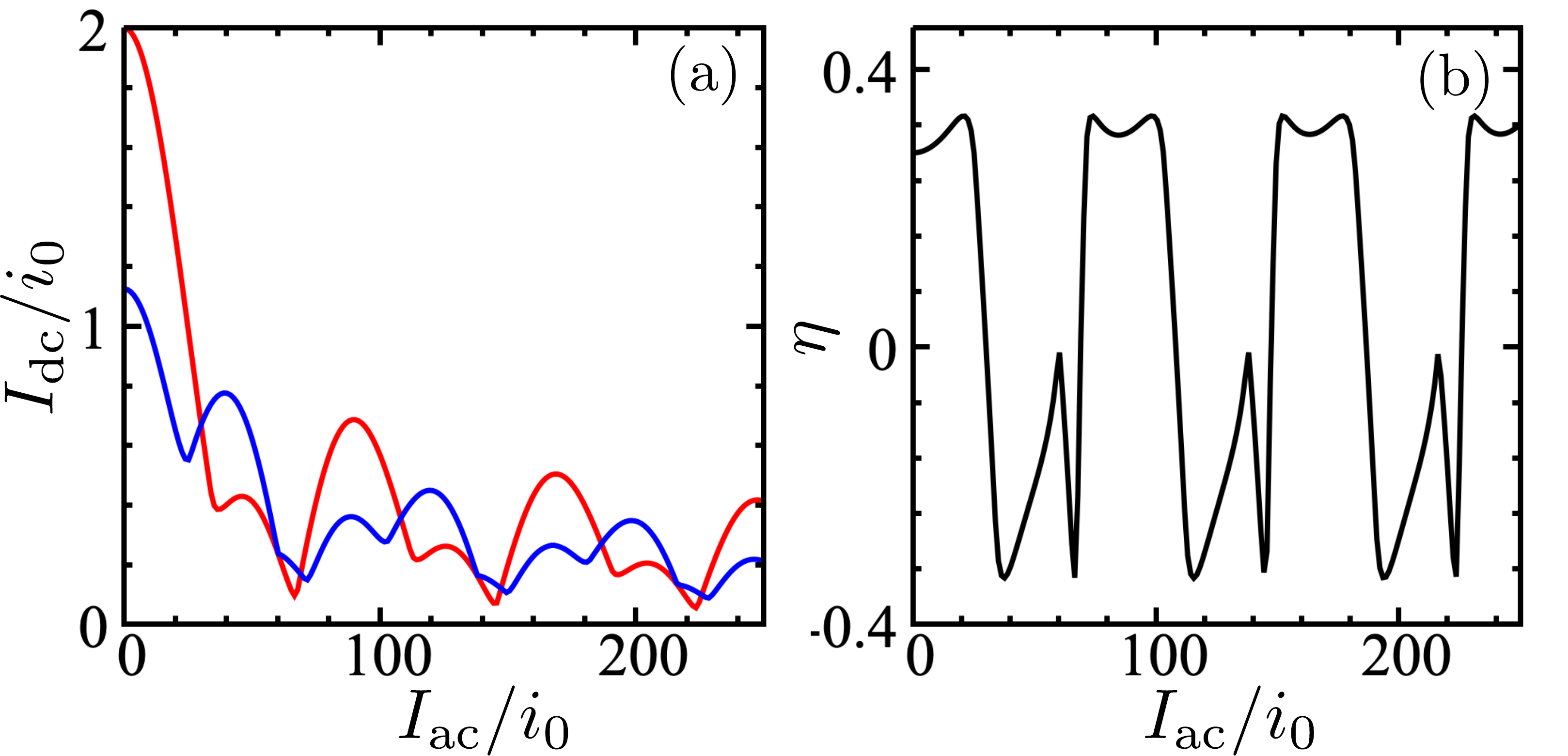}
\end{center}
\caption{(a) Critical current in the fast-driving limit in the positive (red) and negative (blue) directions. (b) The corresponding effective diode efficiency. For both panels we used Eq.~(\ref{eq:idc}) to calculate the critical currents.
\label{longRenageFast}}
\end{figure}

Figure~\ref{dissipated_power} shows the dissipated power in the slow and fast regimes as considered in Figs.~1(b) and 3(a) in the main text, respectively.
We used a logarithmic scale to enhance features. In the slow-driving regime, the system shows a region with significantly reduced dissipated power, see Fig.~\ref{dissipated_power}(a). In contrast, the dissipation in the fast-driving regime increases rapidly as $P=I^2_{\rm ac}R/2$ due to the stronger phase drift every cycle, making the zero-resistance features not visible, see Fig.~\ref{dissipated_power}(b).
Finally, in Figs.~\ref{dissipated_power}(c,d) we show line cuts of the dissipated power for $I_{\rm dc}=0$, comparing the numerically calculated dissipation (red solid) to the analytic expressions in Eqs.~(\ref{eq:pfast}) and (\ref{eq:pslow}) (blue dashed). Both approximate expressions describe accurately the energy dissipated in these two limits. In the slow-driving regime, the analytic expression starts to deviate from the analytic result close to the boundary of the resistive regime, where corrections to the harmonic approximation of the Josephson potential start to become important.

\begin{figure}[t]
\begin{center}
    \includegraphics[width=1\linewidth]{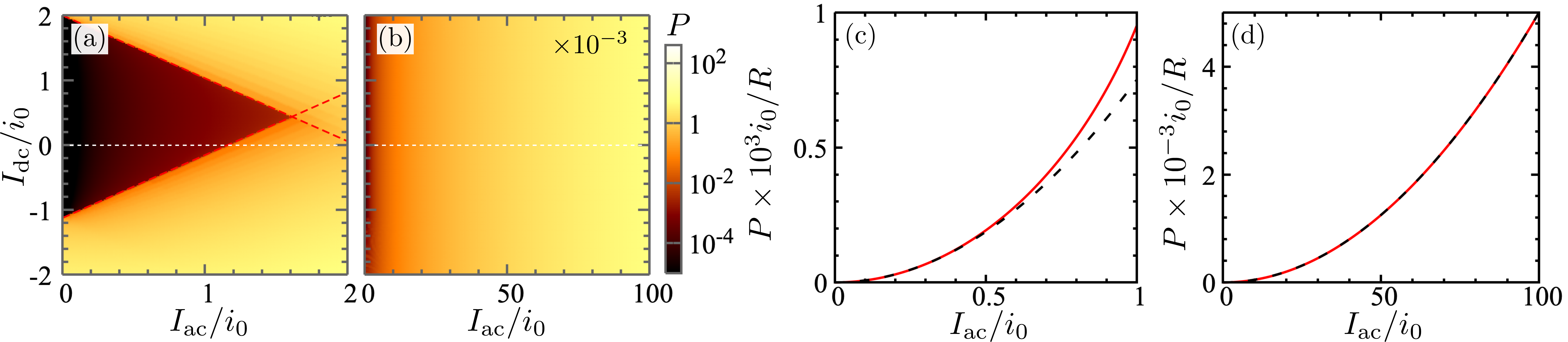}
\end{center}
\caption{Dissipated power. We show numerically calculated results in the slow (a) and fast (b) driving regimes. The corresponding parameters are the same as those used in Figs.~1 and 3, of the main text. Panels (c) and (d) show line cuts for the dissipated power at $I_{\rm dc}=0$ for slow and fast driving, respectively [parameters are the same as in panels (a,b)]. Dashed lines show the analytic results using Eq.~(\ref{eq:pslow}), with the obtained $\alpha=1.29$ for the used current--phase relationship, and Eq.~(\ref{eq:pfast}).
\label{dissipated_power}}
\end{figure}
\end{widetext}

\bibliography{bibliography}

\end{document}